\documentclass[conference]{IEEEtran}
\IEEEoverridecommandlockouts

\usepackage{cite}
\usepackage{amsmath,amssymb,amsfonts}
\usepackage{algorithmic}
\usepackage{graphicx}
\usepackage{textcomp}
\usepackage{url}
\def\BibTeX{{\rm B\kern-.05em{\sc i\kern-.025em b}\kern-.08em
    T\kern-.1667em\lower.7ex\hbox{E}\kern-.125emX}}
\begin{document}

\onecolumn
\textcopyright 2018 IEEE. Personal use of this material is permitted. Permission from IEEE must be obtained
for all other uses, in any current or future media, including reprinting/republishing this material for advertising or promotional purposes, creating new collective works, for resale or redistribution to servers or lists, or reuse of any copyrighted component of this work in other works.
\newline
\newline
The final, published version of this paper is available under:
S. Marksteiner, "Smart Ticket Protection: An Architecture for Cyber-Protecting Physical Tickets Using Digitally Signed Random Pattern Markers,"
\textit{2018 IEEE 20th Conference on Business Informatics (CBI)}, Vienna, Austria, 2018, pp. 110-113. doi:
10.1109/CBI.2018.10055.
URL: \url{https://ieeexplore.ieee.org/document/8453941/}
\twocolumn

\title{Smart Ticket Protection: An Architecture for Cyber-Protecting Physical Tickets Using Digitally Signed Random Pattern Markers
\thanks{This work was partly supported by the Austrian Research Promotion Agency (FFG) within the Austrian security research
program \textit{KIRAS}, grant nb. 845491 (project \textit{Securestamp}), of the Federal Ministry for Transport, Innovation and Technology (BMVIT).}
}

\author{\IEEEauthorblockN{Stefan Marksteiner}
\IEEEauthorblockA{
	\textit{DIGITAL - Institute for Information} \\
	\textit{and Communication Technologies}\\
	\textit{JOANNEUM RESEARCH GmbH}\\
	Graz, Austria\\ 
	Email: stefan.marksteiner@joanneum.at
	}
}

\IEEEspecialpapernotice{(Workshop Paper)}
                
\maketitle

\begin{abstract}

In order to counter forgeries of tickets for public transport or mass events, a method to 
validate them, using printed unique random pattern markers was developed.
These markers themselves are unforgeable by their physically random distribution. To assure their authenticity, however,
they have to be cryptographically protected and equipped with an environment for successful validation, combining physical and cyber security protection.
This paper describes an architecture for cryptographically protecting these markers, which are stored in Aztec codes on
physical tickets, in order to assure that only an
authorized printer can generate a valid Aztec code of such a pattern, thus providing forge
protection in combination with the randomness and uniqueness of the pattern. 
Nevertheless, the choice of the signature algorithm is
heavily constrained by the sizes of the pattern, ticket provider data, metadata and the signature confronted
by the data volume the code hold. Therefore, this paper also defines an example for a signature layout for the proposed
architecture.
This allows for a lightweight ticket validation system that is both physically and cryptographically secured to form a
smart solution for mass access verification for both shorter to longer periods at relatively low cost.  

\end{abstract}


\section{Introduction}
\label{sec:int}
Printed tickets are subject to forgery by (partially organized \cite{ayling2012suppression,Schneider2017}) criminals,
as they may achieve high revenues at black market places \cite{tuffanelli2014super} with a comparably low effort of producing them. To
mitigate this, the research project
\textit{Securestamp}\footnote{http://www.kiras.at/gefoerderte-projekte/detail/d/securestamp/}
has developed a method of applying physically unique, infrared-visible marker pigments on tickets, based on a project partner's patented work 
\cite{Ulrich2014}.
These pigments could be read and should provide a means to assure the authenticity and integrity of said tickets in
conjunction with a digital checking component in form of this marker pigment pattern transformed in an Aztec code.
In order to protect this Aztec code's authenticity and integrity, it has to be cryptographically protected.
This paper outlines the composition of this cryptographic protection and provides an architecture that applies the
former in a setting of printers and readers for ticket production and verification, allowing for the composition of 
physical and cryptographic security to form a lightweight, smart solution for mass access verification.
Based on reading tests with Aztec codes\cite{ISO:24778}, the latter provide space for 704 bytes for actual coding
information. Due to the size of the pigment pattern, 512 bytes are reserved for its Aztec representation and additional
32 bytes are reserved for provider data.
This yields a remaining storage
capability of 160 bytes for encoding protective information within the Aztec code. Therefore, this is the
fundamental working size for elaborating the cryptographic protection of the code.

\section{Related Work}
\label{sec:rel}
Although, in principle, the proposed architecture could be used with any form of unique physical attribute that can 
be encoded (into an arbitrary code), the underlying use case is encoding and protecting the infrared-visible marker
pigments mentioned above into an Aztec code. Therefore, that marking method \cite{Ulrich2014} can be seen as foundation
of the presented work.
Previously known work in this area does, in contrast to the solution described in this paper, not
give comprehensive advice on cryptographically protecting physical ticket features. It either focuses on the hardness
of forging the physical characteristics combined with challenges for verification  \cite{5207652} or does not give any
considerations on cryptographically protecting a code from being copied \cite{6064445}.
Furthermore, previous approaches do not deal with the possibility of stolen ticket printers or
verification devices.

\section{Methodology}
\label{sec:arc}
From a system architectural perspective, the process of issuing and verifying tickets needs protection on three
different levels:

\begin{itemize}
	\item At a physical level, the authenticity and integrity of the marker pigments and Aztec codes have to be assured; 
	\item At device level, secure key management has to take place;
	\item At network level, the key exchange has to be appropriately secured.
\end{itemize}
A means for achieving security at physical level is using \textit{digital signatures} (see Section \ref{sec:cry:sig}). 
The key management needed at device level induces the need for a common format storing and exchanging cryptographic keys.
This can be achieved through the use of \textit{digital certificates} (e.g.  ITU-T X.509\cite{ITU:X509} or IETF RFC
5280\cite{RFC5280}) or a format based on the former. Another possibility is exchanging public key lists, preferably
based on an internationally recognized specification, e.g. the format of the International Union of Railways
(UIC)\footnote{https://railpublickey.uic.org}, which is based on the \textit{Extensible Markup Language
(XML)}\cite{bray1997extensible}.
\begin{figure*}[htbp]
	\centerline{
		\scalebox{0.4}{
			\includegraphics{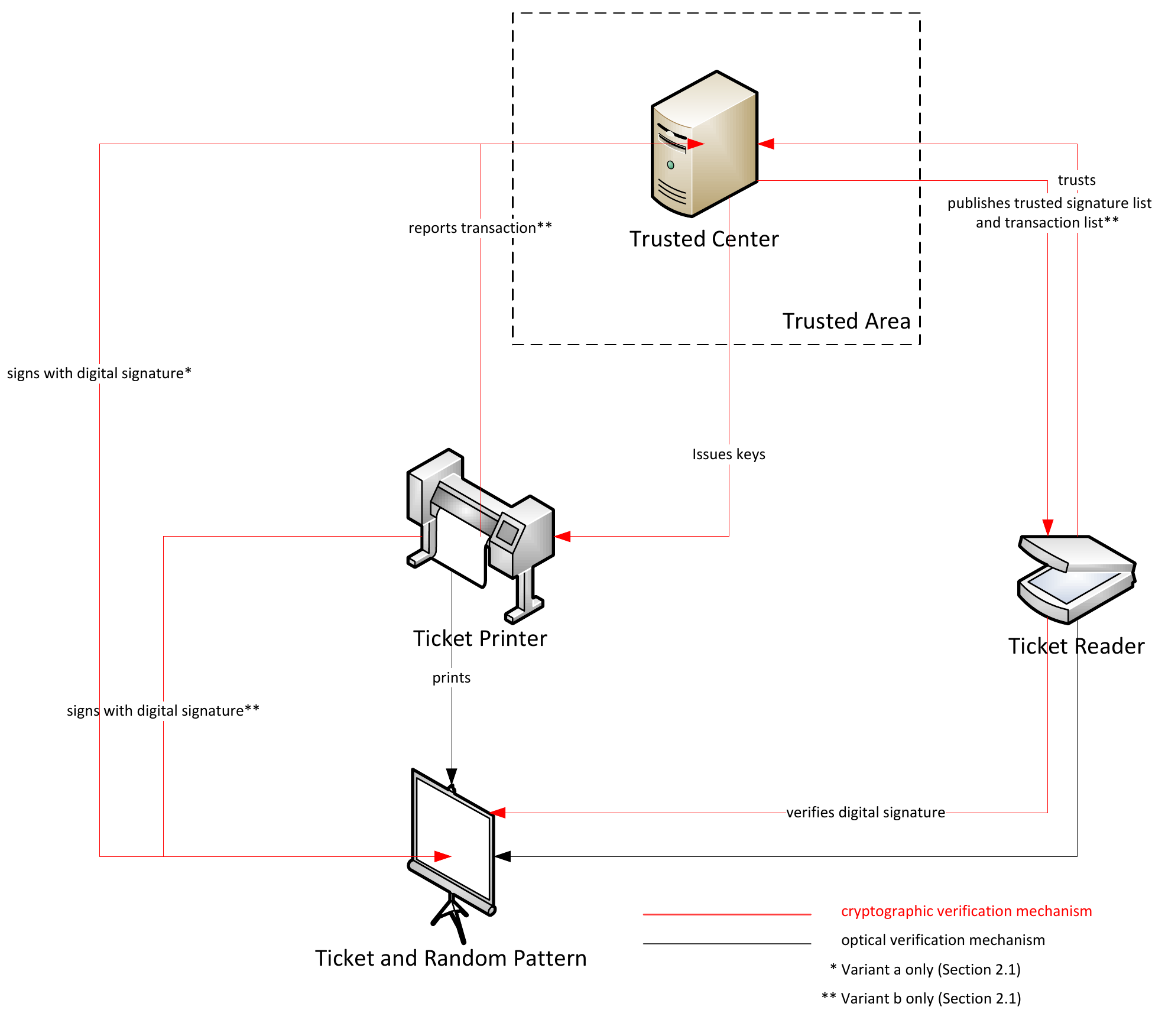}
		}
	}
\caption{Example architecture layout}
\label{fig:arc}
\end{figure*}
Securing the key exchange at network level needs some kind of Public Key Infrastructure (PKI). This can be achieved
using a \textit{Trusted Center (TC)}, where communications between printer or reader and the TC have to occur using
appropriately secure protocols in a configuration set that is also deemed secure (for recommendations see
\cite{Marksteiner:2017:7973855}).
This \textit{Trusted Center} consists of a secure server system under full control of the ticket
provider, securely located in a protected network inside of its data center. This system identifies itself to readers
and printers using an own digital certificate for communication (therefore using separate communication keys),
preinstalled and being trusted a priori on the client devices.
Subsequently (using a secure protocol as mentioned above), readers can request a list of currently valid code signing keys in an according format
(which consists, dependent on the chosen variant, of public keys of the participating TC(s) and/or currently authorized
printers) from the TC. Through this operation, invalidated keys (e.g. of reportedly stolen or compromised printers) are
removed from the trusted keys list and tickets issued by invalidated printers (from the time
invalidation onward) also become invalid (see Section \ref{sec:arc:bak}). All tickets not featuring a successfully
verifiable signature are also regarded invalid. Figure \ref{fig:arc} shows an example for a possible system architecture.

\subsection{The Problem of Backdating}
\label{sec:arc:bak}
The possibility of backdating tickets poses an aggravation to the problem of forging tickets. In principle, tickets
issued by stolen printers can be trivially marked as invalid from the reported time of theft onward. For long-term
tickets, however, the situation is more complicated. A rouge printer with a legitimate but invalidated key may issue
backdated (i.e. before the point of invalidation) but still usable tickets (because of a longer validity period - e.g.
annual tickets), which, per se, are indistinguishable from legitimate tickets issued before the theft. This way, a rouge
printer could issue illegitimate but valid tickets. This can only be prevented if the TC (which must not be stolen or
compromised) performs the signature or the security features (i.e. the signed marker pigment pattern) contain a
timestamp verifying also the issue time. This yields two variants of ticket signing:
\begin{itemize}
  \item The TC performs each signing operation;
  \item Printers issuing tickets report each operation to the TC.  
\end{itemize}
Using the first method, the printer sends the raw information for the signature code (including the marker pigment
pattern) to the TC via a secure channel (in the sense mentioned above), which subsequently signs the ticket
with its own ticket signing key. In scenario, only TCs are in possession of ticket singing keys. Other keys, in
possession of printers and readers, are only dedicated to securing communications between those and the TC(s).

The second method requires issuing printers to transmit the data forming the Aztec code, at least its printer ID, the
ticket ID, the marker pigment pattern, a timestamp and the issued ticket's validity period to the TC. If, in this
scenario, a printer is stolen (and is therefore able to issue arbitrary illegitimate but valid tickets) its ticket
singing key will be rendered invalid by the TC as soon as the theft is discovered, generally invalidating all its issued
tickets.
At the same time, the TC will cease to accept transactions of the printer in question (for certificate-based communications this technically means that the TC will
discontinue to accept this printer's communications certificate). In order to render legit tickets, actually issued
before the theft, still valid, ticket-validating reader devices will be able to request a list of legitimate
transactions from the TC over the secure channel. This can happen in its entirety or, to save memory, only in parts -
e.g. IDs, timestamps and parts or hashes of the respective marker patterns (leaving a certain residual risk of false
positives). For these known transactions, the key of the stolen issuing device will be exceptionally accepted, allowing
for legitimate tickets to retain their validity. This, however, requires all reader devices to be capable of storing all of these transactions for
the remainder of their ticket's validity time. In return, no permanent online connection for the reader devices is
necessary, for the list of transactions can be transferred periodically (e.g. daily). Transactions not reported before
the theft, however, are still invalid and will have to be reissued, which is a neglectable issue if the printer devices
possess permanent online connection.

\subsection{Recommended Signature Schemes}
\label{sec:cry:sig}
Marker pigment patterns and ticket data contain information that has, while having no particular need for secrecy, to be
unforgeable, meaning their authenticity and integrity has to be intact. The latter two fundamental requirements to the
cryptographic protection of the stored code can be met by digital signatures\cite{NIST:2016}.

Digital signatures, like cryptographic functions in general, leave the choice between symmetric and asymmetric
algorithms. Former have the disadvantage that they either:
\begin{itemize}
  \item require a distinct key for each relation between signer (TC or printer) and validator (reader), requiring  
  $O(n^2)$ keys or
  \item require changing all signer keys (i.e. re-installing new ones on all validator devices) in the event of the
  theft of a single validator, as all signer keys have to be known to all validators and are therefore exposed in the
  above event.
\end{itemize}
Asymmetric schemes do not have these disadvantages, because the respective public key is actually intended to be known.
This means that the theft of a validator does not compromise any signer key. Therefore,
the marker pigment patterns and the necessary additional information can be coded directly into the 2D code and be
protected by a digital signature, as this information does not need secrecy but very well needs authenticity and
integrity. For this reason, asymmetric methods are preferable for this use case.
The specifications in \cite{FIPS:186-4} provide standardized schemes, of which the first, the Digital Signature Standard
(DSS) provides three possibilities of generating digital signatures:
\begin{itemize}
	\item \textit{The Rivest-Shamir-Adleman (RSA) algorithm};
	\item \textit{The Digital Signature Algorithm (DSA)};
	\item \textit{The Elliptic Curve Digital Signature Algorithm (ECDSA)}.
\end{itemize}
ISO 14888-3 \cite{ISO:14888} defines twelve standards, partly overlapping with the DSS or variants of DSA or ECDSA. Not variants of these are:
\begin{itemize}
	\item \textit{Variants of the (EC)Schnorr (as (EC-)(F)SDSA)} algorithm;
	\item \textit{The Pointcheval and Vaudeney (PV)} scheme.
\end{itemize}

Of the above, the RSA algorithm is ruled out, as secure key lengths for future applications using this algorithm have
to be more than 4096 bits \cite{ENISA:2014} and the ciphered code length is equal or longer than the key length (in
fact, with a maximum of 132 bytes of space, anything above RSA-1024 does not fit into the signature space). The
private key-enciphered hash value of the digital signature would therefore yield a minimum length of 512 bytes, which exceeds the provided
space. The DSA on the other hand, is recommended for legacy applications only by the \textit{European Union Agency for
Network and Information Security (ENISA)} and is, therefore, also not suitable \cite{ENISA:2014}.
Apart from the exclusion of RSA and DSA algorithms, the relatively short key length suggests \textit{Elliptic Curve
Cryptography (ECC)} algorithms for the present use case. Digital signature generation and verification with ECDSA using
key lengths $>=$224 bits provides an acceptable level of security \cite{NIST:2015}. This also rules out PV scheme, as it
is, despite being easily adaptable to ECC, only standardized for finite fields and also suffers from poor key randomness
\cite{ENISA:2014}. In any case, the hashing algorithm used in the signature should at least be the \textit{Secure Hash
Algorithm 2 (SHA-2)} with an output length $>=$224 bits \cite{ENISA:2014}.
The ECC algorithm combinations fulfilling this criterion also fulfil the length requirements mentioned above. Despite of
this, the ENISA recommends only the algorithms (EC)Schnorr and (EC)KDSA (Korean DSA) for future applications of the ECDSA.
The reason is the weak formal security proof of other ECDSA algorithms. (EC)KDSA, however, cannot be used, for it lacks
of a reference implementation. Schnorr's algorithm \cite{10.1007/0-387-34805-0_22} was patented, but this patent has
expired and also an optimized version of the algorithm (EC-SDSA-opt) has become part of an ISO standard \cite{ISO:14888}.
Therefore, it is preferable for its security proof and also for it can achieve shorter signature sizes than the DSA
\cite{ENISA:2014}. Furthermore, due to its simplicity, it allows for optimizations and it also implements randomized
hashing \cite{ETSI:TS-119312}. Although this algorithm was not well proliferated so far, it got worldwide attention by
\textit{Bitcoin}'s announcement to replacing the ECDSA with Schnorr as its digital signature algorithm in 2017
\cite{Bitcoin:2017}. Apart from the above, all cryptographic functions require a cryptographically secure random number
generator (for ECDSA, for example, according to the recommendations from the German \textit{Bundesamt f\"ur Sicherheit
in der Informationstechnik (BSI)}\cite{BSI:2015} or the \textit{American National Standards Institute's (ANSI)} standard
\textit{X9.62}\cite{ANSI:X962}).

\subsection{Resulting Signature}
\label{sec:cry:res}
\begin{figure}[htbp]
	\centerline{
		\scalebox{0.45}{
			\includegraphics{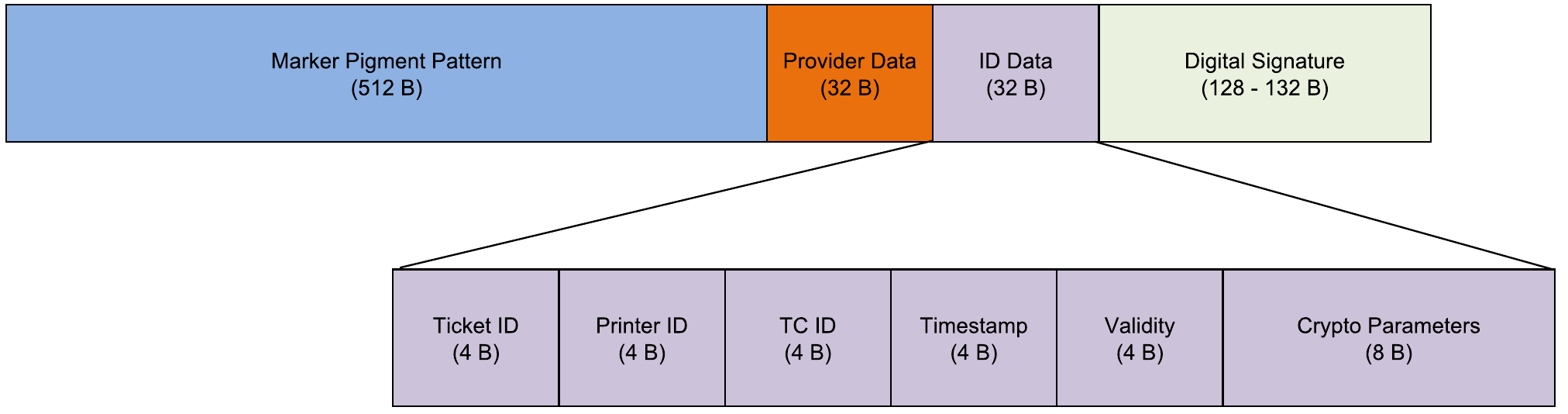}
		}
	}
\caption{Example code layout.}
\label{fig:code}
\end{figure}

The resulting Aztec code should, apart from the marker pigment pattern and the digital signature, contain at least an
identification number (ID) for each the issuing printer, the TC, the ticket provider and the ticket itself, as
well as the ticket's validity time (period or date-of-expiry) and an issue timestamp (Figure \ref{fig:code} shows an exemplary
code layout). Information already contained in the provider data (32B) may be omitted. For better interoperability, the
code may hold information about cryptographic parameters (e.g. used elliptic curve, hash algorithm, etc.). The digital
signature must protect all code components (except the signature itself).

\section{Conclusion and Outlook}
This work demonstrated that \textit{Elliptic Curve Cryptography (ECC)} provides a means to securely embed a
physically unique marker pigment pattern into Aztec codes. It further outlined an architecture that allows for
successful (part-offline) issuing and cryptographic verification of tickets and, especially, to overcome the problem of
distinguishing legitimate from illegitimate long-term tickets, validly issued by legit, but stolen or corrupted printers
using transaction tracking or digital signing in a \textit{Trusted Center (TC)}. Future work will include implementing
and testing the presented solution and further developing the latter to forge a market-ready product.

\section*{Acknowledgment}
The author wants to thank his fellow colleague Martin Winter for his support in this work.

\bibliographystyle{IEEEtran} 
\bibliography{IEEEabrv,literature}

\end{document}